\documentclass[11pt]{article}

\usepackage[preprint]{acl}

\usepackage{times}
\usepackage{latexsym}
\usepackage{booktabs}

\usepackage[T1]{fontenc}

\usepackage[utf8]{inputenc}

\usepackage{microtype}

\usepackage{inconsolata}

\usepackage{graphicx}

\title{DuplexOmni: Real-Time Listening, Seeing, Thinking, and Speaking for Full-Duplex Interaction}

\author{
\textbf{Muye Huang}\textsuperscript{1,2},
\textbf{Lingling Zhang}\textsuperscript{1,2,\textdagger},
\textbf{Xingyu Yu}\textsuperscript{3},
\textbf{Lei Shi}\textsuperscript{4},
\textbf{Zhanyu Ma}\textsuperscript{4}, \\
\textbf{Jun Xu}\textsuperscript{4,\textdagger},
\textbf{Jiuchong Gao}\textsuperscript{4},
\textbf{Jinghua Hao}\textsuperscript{4},
\textbf{Renqing He}\textsuperscript{4},
\textbf{Jun Liu}\textsuperscript{1,2} \\
$^1$School of Computer Science and Technology, Xi'an Jiaotong University \\
$^2$MOE KLNN Lab, Xi'an Jiaotong University \\ 
$^3$School of Software and Microelectronics, Peking University 
$^4$Meituan \\
\texttt{huangmuye@stu.xjtu.edu.cn} 
$^\dagger$Corresponding authors.
}

\begin{document}
\maketitle
\begin{abstract}
Human interaction is continuous, multimodal, and full-duplex by nature. Although recent omni models have made substantial progress in unified speech, vision, and text modeling, combining seamless real-time interaction with complex reasoning and tool use remains challenging. We present DuplexOmni, a method for real-time multimodal full-duplex interaction. DuplexOmni separates model capability into an interaction layer and a thinking layer, which collaborate asynchronously in parallel. The interaction layer is implemented by the DuplexOmni model, an end-to-end system that processes streaming audio and video inputs while generating text and speech responses in real time. The thinking layer is a pluggable module that provides complex reasoning and tool-use capabilities. To support this method, we further develop a Writer-Director pipeline for constructing continuous-interaction training data. Experiments show that DuplexOmni achieves strong performance on multiple public benchmarks and exhibits natural full-duplex interaction ability. 
\end{abstract}

\begin{figure}[ht]
    \centering
    \includegraphics[width=1\linewidth]{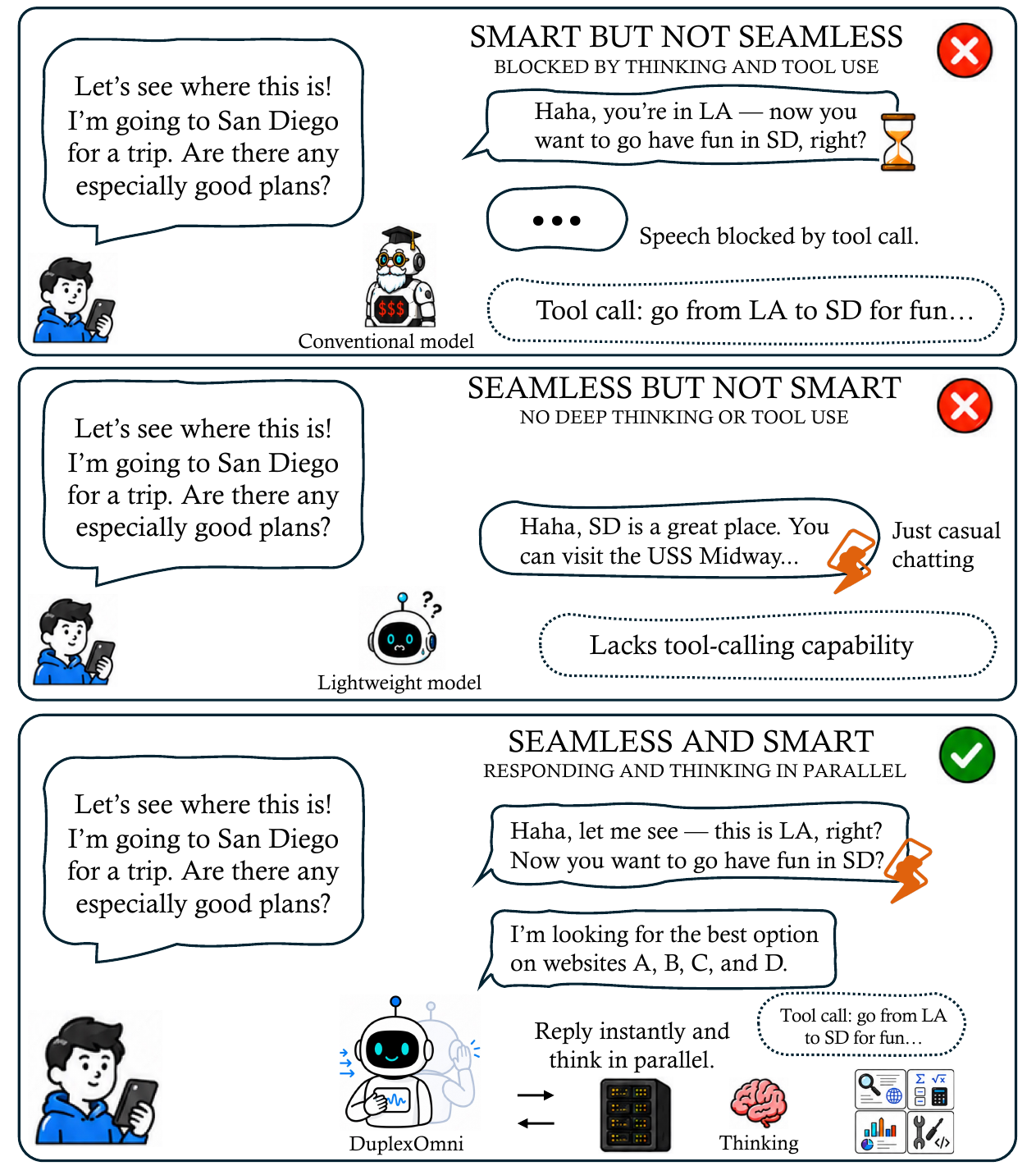}
    \caption{DuplexOmni keeps real-time interaction continuous by decoupling instant responses from asynchronous thinking and tool use.}
    \label{fig:placeholder}
\end{figure}

\section{Introduction}

Human interaction is naturally seamless, multimodal, and full-duplex. During communication, people listen to speech, observe the environment, and understand context, while continuing to receive new information as they speak. New information may change subsequent expression or lead to interaction events such as pauses. Therefore, this kind of interaction requires a model to express in real time during continuous perception and to keep updating its understanding during expression.

Recent omni models \cite{DBLP:journals/corr/abs-2410-21276, DBLP:journals/corr/abs-2503-20215, yao2024minicpm, DBLP:journals/corr/abs-2403-05530} have advanced unified modeling of speech, vision, and text. These models have demonstrated real-time multimodal understanding and generation. However, existing models still struggle to provide seamless user interaction. As shown in Fig.~\ref{fig:placeholder}, when interaction involves deep thinking or tool use, the model often pauses the ongoing dialogue. It then waits for reasoning or tool results before continuing its response. This causes clear interruptions. Some methods build lightweight real-time models by reducing model size and simplifying reasoning. However, this weakens or even removes the ability to handle complex tasks. The root cause is that listening, speaking, reasoning, and acting are usually handled within a single serial pipeline. Reasoning and tool use can interrupt real-time interaction, while shortening the reasoning process limits the model’s capabilities. Seamless multimodal interaction therefore requires independent execution spaces for real-time interaction and deep reasoning, as well as continuous collaboration throughout the dialogue.

To address this contradiction, we propose DuplexOmni, a method for real-time multimodal interaction that enables seeing, listening, speaking, thinking, and acting to proceed in parallel. Inspired by fast and slow thinking in human cognition, DuplexOmni divides model capabilities into a decoupled interaction layer and thinking layer, and lets them collaborate asynchronously. The interaction layer is handled by the DuplexOmni model, an end-to-end full-duplex model that continuously receives real-time speech, video, and dialogue history, while producing streaming text and speech. The thinking layer provides pluggable deep cognitive capability, and can be instantiated with large language models or tool agents according to the task. The interaction layer maintains low-latency perception and expression, and requests background thinking from the thinking layer when needed. Results returned by the thinking layer are progressively received, organized, and incorporated by the interaction layer into subsequent responses, keeping real-time interaction continuous while giving deep thinking independent computation time.

To realize DuplexOmni, we develop the model, collaboration mechanism, data construction, and experimental validation. First, we train the DuplexOmni model as the interaction layer, enabling an end-to-end model to jointly handle real-time speech, video, dialogue state, text output, and speech output. Second, we design asynchronous collaboration between the interaction layer and the thinking layer, allowing the interaction layer to invoke external thinking capability and integrate returned results during ongoing dialogue. Third, we build a Writer-Director data pipeline for continuous interaction, covering overlapping speech, silence, user supplements, interruptions, delayed thinking feedback, and complex task progression. Experiments show that DuplexOmni achieves advanced performance on multiple benchmarks and exhibits natural full-duplex interaction abilities. We will release the model weights, training data, and training and inference implementation.

Our main contributions are as follows:

\begin{itemize}
    \item We propose DuplexOmni, a continuous multimodal interaction method that decouples the interaction layer from the thinking layer, enabling low-latency interaction and pluggable deep thinking to collaborate asynchronously.

    \item We build a Writer-Director data pipeline, train the DuplexOmni model with the resulting data, and provide a new open-source solution for constructing such interaction data.

    \item We validate DuplexOmni on multiple public benchmarks, and will release the full training and inference process to support future research.
\end{itemize}

\begin{figure*}[ht]
    \centering
    \includegraphics[width=1\linewidth]{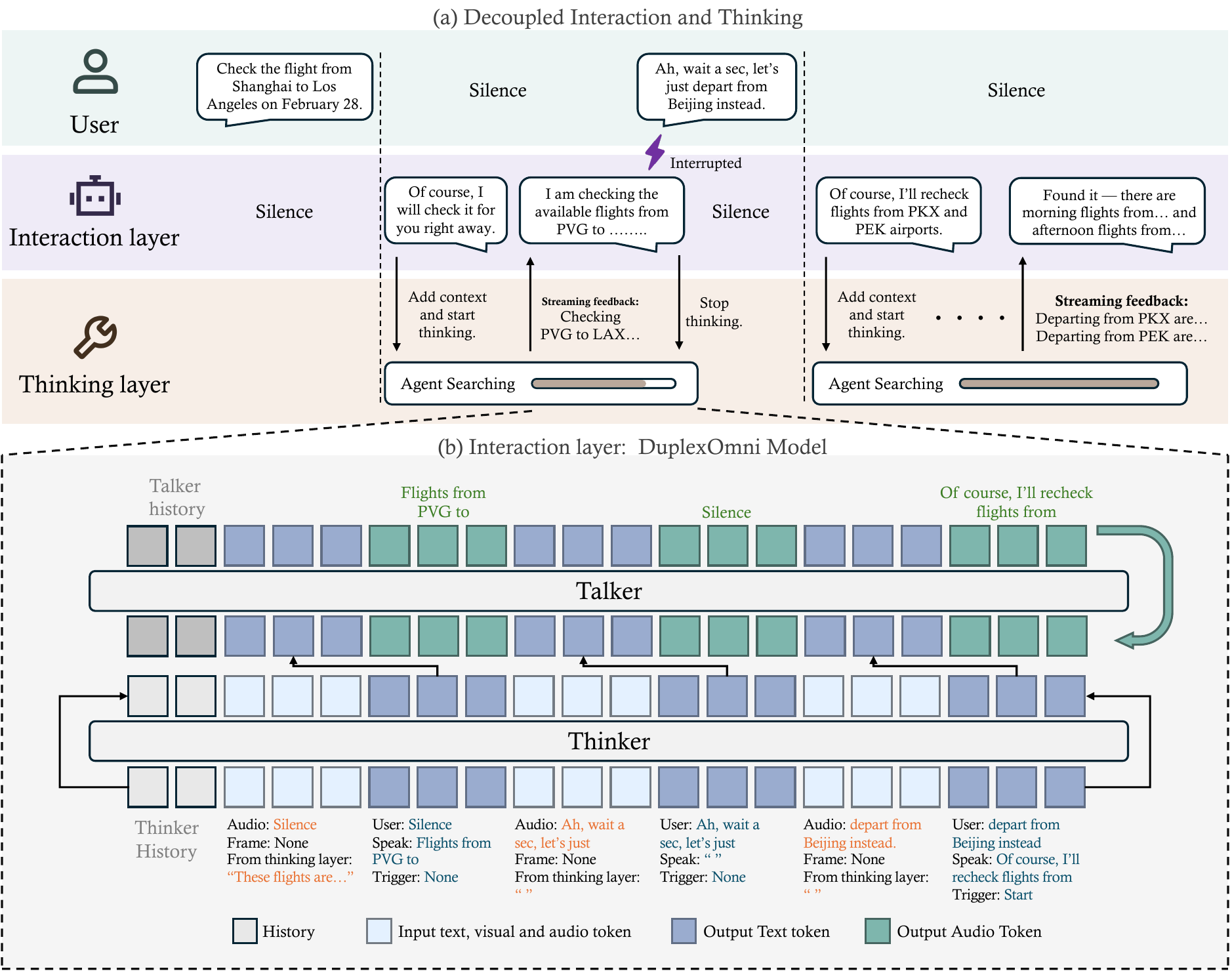}
\caption{
Overview of DuplexOmni. 
(a) The interaction layer conducts real-time dialogue while asynchronously collaborating with the thinking layer. 
(b) The DuplexOmni model implements the interaction layer with time-sliced Thinker-Talker inference over streaming inputs and thinking feedback. MTP and Code2Wav are omitted for clarity.
}

    \label{fig:overview}

\end{figure*}

\section{Related work}
\subsection{Omni Models}

Recent years, substantial progress in omni models \cite{DBLP:journals/corr/abs-2512-20156, DBLP:conf/acl/Chen0YLLXN00L0025}. By introducing end-to-end audio-visual understanding and generation, these models can process multiple forms of information in a way closer to human interaction. Models such as GPT-4o \cite{DBLP:journals/corr/abs-2410-21276}, Gemini \cite{DBLP:journals/corr/abs-2312-11805}, Qwen2.5-Omni \cite{DBLP:journals/corr/abs-2509-17765}, Qwen3-Omni \cite{DBLP:journals/corr/abs-2509-17765}, MiniCPM-o \cite{yao2024minicpm}, Baichuan-Omni \cite{DBLP:journals/corr/abs-2501-15368}, and Kimi-Audio \cite{DBLP:journals/corr/abs-2504-18425} have demonstrated strong cross-modal understanding and dialogue capabilities. The Qwen series unifies speech, vision, and text capabilities within an end-to-end omni architecture; MiniCPM-o emphasizes efficient multimodal interaction in on-device or lightweight settings. Despite significantly expanding the perceptual and generative boundaries of large models, most of these models still organize interaction in a request-driven manner: user input is collected, the model performs understanding and generation, and then a response is returned. This mode is suitable for single-turn or multi-turn task processing, but cannot handle user interaction in real time. Therefore, omni-modal models provide a foundation for unified multimodal capability, but do not directly address coordination in real-time continuous interaction.

\subsection{Duplex Interaction}

Other studies focus on duplex interaction, where a model can receive user input and generate responses simultaneously in real time. Unlike traditional simplex models, duplex interaction \cite{DBLP:conf/aaai/MaSDC0WW025, nguyen2022generativespokendialoguelanguage, DBLP:journals/corr/abs-2405-19487, DBLP:journals/corr/abs-2509-06502, DBLP:journals/corr/abs-2506-01268, DBLP:journals/corr/abs-2603-09180} requires the model to keep listening to user speech while speaking, and to handle natural interaction phenomena such as overlapping speech and interjections. Moshi \cite{DBLP:journals/corr/abs-2410-00037} achieves low-latency full-duplex speech dialogue by jointly modeling user speech, assistant text, and assistant speech. OmniFlatten \cite{DBLP:conf/acl/ZhangCDCWZLYTDZ25} flattens speech and text inputs and outputs into a unified GPT sequence to model seamless voice conversation in an end-to-end manner. MiniCPM-o \cite{yao2024minicpm} models time-aligned multimodal streams, enabling the model to see, listen, and speak on a unified timeline. Mini-Omni-Reasoner \cite{DBLP:journals/corr/abs-2508-15827}, and dGSLM \cite{nguyen2022generativespokendialoguelanguage} also advance streaming speech modeling and parallel listening and speaking \cite{DBLP:journals/corr/abs-2505-17060, DBLP:conf/emnlp/ZhangLZZWZQ23, DBLP:conf/acl/HeldZSLRY25, DBLP:journals/corr/abs-2512-23808} from different perspectives. These methods move speech interaction from strict turn-taking toward more natural continuous dialogue. However, when dialogue requires longer reasoning, interaction between the model and the user is still interrupted.

\section{Method}

We propose DuplexOmni, a method for multimodal full-duplex interaction. As shown in Figure~\ref{fig:overview} DuplexOmni divides model capabilities into an interaction layer and a thinking layer, and connects real-time interaction with deep thinking through asynchronous parallel collaboration. The interaction layer is implemented by the DuplexOmni model, which continuously receives speech, video, dialogue state, and results returned from the thinking layer, organizes the current context, infers user intent, and generates text and speech responses in real time. The thinking layer is a pluggable layer that can be instantiated with MLLMs or tool agents to perform complex reasoning, tool use, and task planning. To train and deploy this form of interaction, we build a Writer-Director data pipeline that generates continuous-interaction data and simulates realistic dialogue scenarios for training the full-duplex interaction model. In the following sections, we detail the hierarchical structure of DuplexOmni, the Writer-Director data pipeline, and real-time inference for low-latency deployment.

\subsection{DuplexOmni}
DuplexOmni adopts a two-layer structure with an interaction layer and a thinking layer. The interaction layer handles real-time interaction: it reads audio, video, and intermediate results from the thinking layer, organizes responses to the user, controls the dialogue rhythm, and requests assistance from the thinking layer when needed. The thinking layer provides additional cognitive capability through a pluggable interface, running complex reasoning and tool use in the background. We first describe the decoupled structure, then the DuplexOmni model that implements the interaction layer, and finally the running strategy of the thinking layer.

\subsubsection{Decoupled Interaction and Thinking}
Decoupled Interaction and Thinking centers on thinking requests initiated by the interaction layer and streaming results returned by the thinking layer. The interaction layer handles real-time dialogue, while the thinking layer performs background reasoning. When the current interaction requires external assistance, the interaction layer passes the user context to the thinking layer, including dialogue text, video information, and task state, and issues a thinking request. This request does not block real-time interaction. The interaction layer continues to receive user input, control the dialogue rhythm, and organize immediate responses.

After receiving a request, the thinking layer runs asynchronously with the interaction layer. It can be a strong LLM or a task-specific agent. During streaming output, the thinking layer wraps intermediate results with special control tokens and continuously returns them to the interaction layer. The interaction layer organizes these responses into text and speech replies suitable for the current context. When the interaction layer determines that conditions have changed or that the current dialogue no longer needs external assistance, it stops the streaming output of the thinking layer through special control tokens. This design enables DuplexOmni to continuously serve users in low-latency interaction while incorporating complex reasoning, tool use, or specialized agent capabilities on demand.

\subsubsection{DuplexOmni Model}

\paragraph{Time-Sliced Full-Duplex Modeling}
DuplexOmni model realizes full-duplex interaction through time-sliced autoregressive inference. Similar to Moshi \cite{DBLP:journals/corr/abs-2410-00037} and MiniCPM-o \cite{yao2024minicpm}, we divide continuous interaction into fixed 480\,ms slices. At slice $t$, the model consumes the dialogue history, the intermediate results returned by the thinking layer, and the inputs from slice $t-1$. For audio, this input is the 480\,ms speech segment in that slice. For video, it is the visual frame sampled from the same slice.

Each slice produces three outputs: a thinking-control signal, a semantic interpretation of the latest user input, and the Assistant's text and 480\,ms speech response. This design allows the model to continue receiving new user inputs and new thinking results while generating responses, thereby achieving full-duplex interaction with a response granularity of 480\,ms.

\paragraph{Model Architecture}
DuplexOmni model follows the Thinker-Talker speech generation structure in the Qwen-Omni \cite{DBLP:journals/corr/abs-2407-10759, DBLP:journals/corr/abs-2509-17765, DBLP:journals/corr/abs-2604-15804} family. The Thinker is the internal MLLM backbone that processes the current context and generates Assistant text tokens. The Talker converts the generated linguistic states into streaming speech. For slice $t$, the Thinker produces an Assistant token sequence together with the aligned token embeddings and hidden states. We denote them by $E_t$ and $H_t$, respectively. These two sequences are projected into the Talker conditioning space and added row-wise:
\[
c_{t,\ell} = f_{\mathrm{text}}(e_{t,\ell}) + f_{\mathrm{hidden}}(h_{t,\ell}), \qquad \ell = 1, \dots, m_t,
\]
where $m_t$ is the number of Assistant tokens in the current slice. We denote the resulting conditioning sequence by
\[
C_t = (c_{t,1}, \dots, c_{t,m_t}).
\]

The Talker does not generate speech from $C_t$ alone. It preserves both the historical conditioning sequences and the historical speech codec history. Let $Q_i$ denote the complete residual vector quantization (RVQ) codec sequence generated in slice $i$. If slice $i$ contains $L_i$ speech frames, then
\[
Q_i = (q_{i,1}, \dots, q_{i,L_i}),
\]
where each frame is represented by
\[
q_{i,j} = (q^0_{i,j}, q^1_{i,j}, \dots, q^{K-1}_{i,j}).
\]
For each frame, the Talker constructs a summed codec embedding
\[
r_{i,j} = u_0(q^0_{i,j}) + \sum_{k=1}^{K-1} u_k(q^k_{i,j}),
\]
and we denote the codec embedding sequence of slice $i$ by
\[
R_i = (r_{i,1}, \dots, r_{i,L_i}).
\]

At the beginning of slice $t$, the embedding prefix of the Talker is
\[
P_t = \bigl[(C_i, b_{\mathrm{BOS}}, R_i, b_{\mathrm{EOS}})_{i=1}^{t-1},\; C_t,\; b_{\mathrm{BOS}}\bigr].
\]

\begin{figure*}[ht]
\vspace{-0.5em}

    \centering
    \includegraphics[width=1\linewidth]{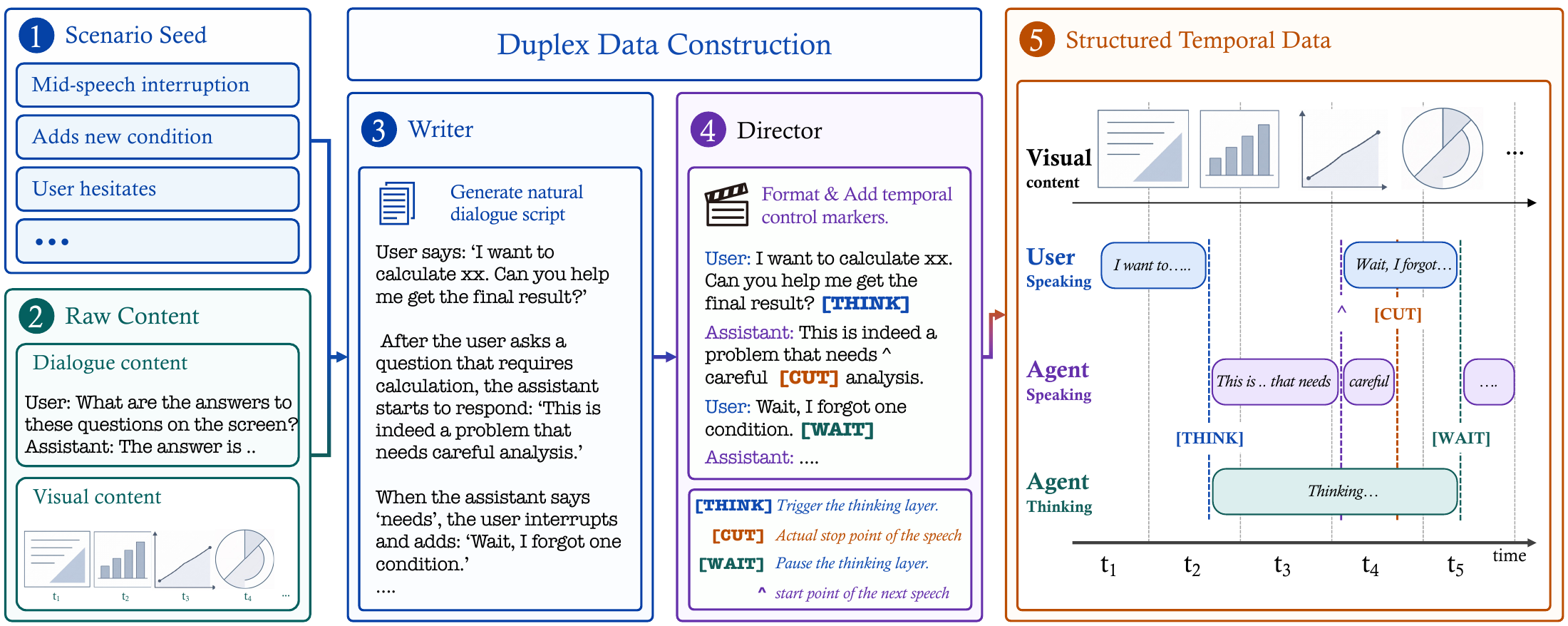}
    \caption{Data pipeline for DuplexOmni. Scenario seeds and raw content are converted into temporally annotated full-duplex dialogues; representative control tokens are shown, with the full token set detailed in the Appendix.}
    \label{datapipeline}
\end{figure*}

This expression matches the actual organization of the prompt. Each historical slice contributes one conditioning segment followed by one codec segment wrapped by a single codec BOS and EOS. The current slice contributes its conditioning sequence and one codec BOS before speech generation starts. Inside the current slice, the model maintains both a token channel and an embedding channel. Conditioning positions are filled with a dedicated pad id on the token channel, while historical codec spans keep their true codec tokens. On the embedding channel, the model writes the conditioning projections and the corresponding summed codec embeddings.

Given the current prefix, the Talker autoregressively predicts the layer-0 codec token of the current speech frame:
\[
q^0_{t,j} = \mathrm{Talker}(P_t, q^0_{t,<j}).
\]
For each non-EOS layer-0 token, an MTP \cite{DBLP:journals/corr/abs-2509-17765} module predicts the residual codebooks conditioned on the current Talker hidden state and the predicted layer-0 token, producing the complete RVQ code $q_{t,j}$ and its summed embedding $r_{t,j}$. The newly generated layer-0 token and summed embedding are then appended to the current prefix and used for the next frame. Therefore, within a slice, the Talker repeatedly appends newly generated codec content, rather than repeatedly inserting new BOS or EOS markers.

Each slice generates six codec frames, corresponding to 480\,ms of speech, and the resulting RVQ sequence is decoded by Code2Wav into waveform samples. After generation, the current conditioning sequence $C_t$ and codec sequence $Q_t$ are written into history and reused to construct the Talker prefix of the next slice. In this way, DuplexOmni model remains autoregressive at the token level, continuous across adjacent time slices, and full-duplex at the interaction level.

\subsection{Data Construction}

To train the full-duplex interaction ability of DuplexOmni, the training data must describe continuous interaction behaviors, such as interruption, backchanneling, waiting, background reasoning, and delayed information integration. Existing multi-turn dialogue data is mostly turn-based. It only records user and assistant utterances, and lacks temporal information. Therefore, it cannot describe when the model should speak, stop, wait, trigger background thinking, or use returned information.

We address this by dividing data construction into two stages: \textit{scenario and content construction} and \textit{Writer-Director temporal annotation}. The first stage decides what the dialogue is about and how it happens. The second stage converts natural dialogue into structured samples with explicit temporal causality. Since content and temporal annotation are decoupled, new tasks only require new content data and scenario configurations. The downstream pipeline remains unchanged.

\subsubsection{Scenario and Content Construction}

This stage builds content-scenario pairs for training. Each sample contains raw content and a scenario seed: the raw content provides the task semantics, while the scenario seed specifies how the interaction unfolds, such as who initiates the dialogue, whether the user may revise or add information during the exchange, and how the assistant should respond when background reasoning or turn-taking events occur. We sample scenario combinations from configuration files, filter invalid or conflicting settings, and rewrite raw content into a speech-friendly form by removing written-style expressions, special symbols, and overly long sentences.

\subsubsection{Writer-Director Data Pipeline}

This stage converts content-scenario pairs into structured temporal data for full-duplex training. The Writer generates a natural dialogue script that follows the scenario. The Director adds temporal control signals.

The Writer generates a natural language script from the raw content and the scenario seed. The script preserves the task semantics and reflects the target interaction pattern. For example, in a math reasoning task, the Writer may arrange the user to ask a question, add missing conditions midway, interrupt the assistant, or insert new information during assistant reasoning. For assistant-initiated scenarios, the Writer adjusts the opening and speaking order so that the dialogue naturally starts from the assistant.

The output of the Writer is still a human-readable linear script. For example: 
\begin{quote}
\small

User says: `I want to calculate xx. Can you help me get the final result?' After the user asks a question that requires calculation, the assistant starts to respond: `This is indeed a problem that needs careful analysis.' When the assistant says `needs', the user interrupts and adds: `Wait, I forgot one condition.'
\end{quote}

This script describes how the dialogue happens. It does not handle control tokens or the audio timeline.

The Director converts this script into a structured sample with temporal control signals. The full definitions of special text tags are shown in Appendix. For the above script, the annotated sample can be written as:

\begin{quote}
\small
User: I want to calculate xx. Can you help me get the final result? \texttt{[THINK]}

Assistant: This is indeed a problem that needs\texttt{\string^} careful\texttt{[CUT]} analysis.

User: Wait, I forgot one condition. \texttt{[WAIT]}
\end{quote}

Here, \texttt{[THINK]} triggers background reasoning. \texttt{\string^} marks the time point where the user speech stream starts. \texttt{[CUT]} marks the actual stopping point of the interrupted assistant speech. \texttt{[WAIT]} means that the user adds a new condition, so the background reasoning should pause or revise. The Director does not change the dialogue content. It adds trainable temporal causality.

Finally, we apply consistency checks to the Director output. We filter samples with conflicting control tokens, missing thinking triggers, returned information without causal source, or unreasonable interruption points. After cleaning, we synthesize speech with TTS and segment the audio into fixed-size chunks, so that the resulting data follows the same time-sliced format used by DuplexOmni model training and inference. The processing details differ between fully synthetic text-to-speech data and dialogue data sampled from real recordings; we leave this discussion to the Appendix. Through this pipeline, DuplexOmni learns when to listen, speak, wait, interrupt, request background thinking, and integrate delayed information.

\subsection{Real-Time Duplex Inference}
To enable real-time duplex inference, DuplexOmni uses $\mathrm{RTF}<1$ as the latency target for speech generation. RTF is the ratio between generation time and audio duration. A value below 1 means that each audio chunk can be generated before it finishes playing. DuplexOmni decouples Thinker-based text generation from Talker-based speech generation and runs them as an asynchronous pipeline. This avoids serial latency accumulation across continuous chunks. For the Talker path, we apply KV-cache-based incremental decoding and graph execution optimization to autoregressive codec decoding and MTP computation. These optimizations reduce redundant computation and scheduling overhead and support stable low-latency speech output.

\section{Experiments}

\begin{table*}[ht]
\vspace{-0.5em}
  \centering
  \setlength{\tabcolsep}{3pt}
  \begin{tabular*}{\textwidth}{@{\extracolsep{\fill}}lccccc}
    \toprule
    Model &
    \begin{tabular}[c]{@{}c@{}}DuplexBench\\ToR (\%)\end{tabular} &
    \begin{tabular}[c]{@{}c@{}}Big Bench\\Audio (\%)\end{tabular} &
    \begin{tabular}[c]{@{}c@{}}Daily-Omni\\(\%)\end{tabular} &
    \begin{tabular}[c]{@{}c@{}}LibriSpeech\\WER\end{tabular} &
    \begin{tabular}[c]{@{}c@{}}Latency\\(s)\end{tabular} \\
    \midrule
    DuplexOmni & \textbf{72.6} & \textbf{77.2} & 53.8 & 0.1192 & 0.506 \\
    MiniCPM-o 4.5 & 36.3 & 36.0 & 80.2 & -- & \textbf{0.502} \\
    Doubao & 27.8 & 47.9 & -- & -- & 1.82 \\
    Qwen3-Omni-Realtime-Flash & 25.2 & 32.1 & 76.2 & 0.013 & 1.28 \\
    Qwen3.5-Omni-Realtime-Flash & 26.4 & 35.0 & \textbf{81.8} & \textbf{0.011} & 1.25 \\
    Gemini-3.1-Flash-Live & 24.1 & 57.9 & -- & 0.041 & 2.57 \\
    Gemini-3.1-Flash-Lite & -- & 58.9 & 59.0 & 0.038 & -- \\
    \bottomrule
  \end{tabular*}
  \caption{Interaction-layer comparison with real-time full-duplex and speech-to-speech baselines. All models with real-time capabilities are evaluated in real-time mode, and Daily-Omni is evaluated under streaming mode.}
  \label{tab:interaction-layer-comparison}
  \vspace{-1em}
\end{table*}

\subsection{Settings}

We compare DuplexOmni with recent real-time omni models and speech-to-speech systems, including MiniCPM-o, Doubao, Qwen-Omni realtime variants, and Gemini live variants. All models are evaluated under their streaming or realtime settings; proprietary models are evaluated through official APIs, and open-source models use public weights with officially recommended inference configurations. We evaluate models on Full DuplexBench (DuplexBench) v1.5 \cite{DBLP:conf/asru/LinLLWALL25}, Big Bench Audio \cite{srivastava2022beyond, suzgun2022challenging}, Daily-Omni \cite{DBLP:journals/corr/abs-2505-17862}, and LibriSpeech \cite{DBLP:conf/icassp/PanayotovCPK15} WER. Full DuplexBench measures full-duplex interaction quality, Big Bench Audio measures streaming audio understanding, and Daily-Omni measures general omni capability. For models with ASR capability, we report LibriSpeech as a reference for speech recognition quality. We also report latency as a measure of real-time response delay. Unless otherwise specified in ablation experiments, DuplexOmni uses the full configuration, where the DuplexOmni model serves as the interaction layer and Gemini-3.1-Flash-Lite \cite{DBLP:journals/corr/abs-2403-05530} serves as the thinking layer.

\subsection{Data}

The training data construction consists of two parts: scenario seeds and dialogue content. We build about 620K scenario seeds to define the distribution of continuous interaction settings. These seeds cover five types of interaction scenarios and annotate attributes such as interaction type, required events, and user style, so that the resulting data spans full-duplex interactions with different densities, rhythms, and levels of complexity.

The dialogue content contains about 3.02M raw conversations, including 10K video-call conversations. The text-dialogue portion includes about 1.486M user-initiated conversations and 1.528M system-initiated conversations, sourced from UltraChat \cite{ding2023enhancing}, WildChat \cite{zhao2024wildchat1mchatgptinteraction} , BELLE \cite{DBLP:journals/corr/abs-2304-07854}, COIG \cite{zhang2023chineseopeninstructiongeneralist}, no-robots \cite{no_robots}, OASST2 \cite{kpf2023openassistantconversationsdemocratizing}. The overall language distribution is mainly Chinese, about 70\%, with about 30\% English and a small amount of other languages. All data construction stages use Qwen3.5-397B-A27B \cite{DBLP:journals/corr/abs-2505-09388} as the generation and annotation model. Speech training data is synthesized with Qwen3-TTS \cite{DBLP:journals/corr/abs-2601-15621} and converted into codec tokens with the Mimi \cite{DBLP:journals/corr/abs-2410-00037} encoder.

\subsection{Training}
DuplexOmni treats time-sliced full-duplex interaction as continuous multi-turn dialogue. Initialized from Qwen3-Omni, it is trained with two-stage SFT: the first stage builds basic full-duplex listening and speaking ability on large-scale speech interaction data, while the second stage further improves complex interaction and video-call performance using high-quality interaction and video-call data.

In each stage, we alternately optimize the Thinker and the Talker. When training the Thinker, we freeze the Talker and compute only the cross-entropy loss of the Thinker. When training the Talker, we freeze the Thinker and compute only the cross-entropy loss of the Talker MoE and MTP modules. The data ratio between the two optimization parts is 1:1. The learning rate is 1e-5 for the Thinker and 1e-4 for the Talker, with a batch size of 128. Training is conducted with Megatron-swift-3.12 \cite{zhao2024swiftascalablelightweightinfrastructure} on 128 Nvidia H20 GPUs.

\subsection{Performance Comparison}

Table~\ref{tab:interaction-layer-comparison} reports the main comparison between DuplexOmni and recent realtime omni models or speech-to-speech systems. The results show three main observations:

\paragraph{Strong full-duplex interaction.} DuplexOmni achieves 72.6\% ToR on Full DuplexBench, substantially outperforming all realtime baselines, while keeping a response latency of 0.506s. This shows that the DuplexOmni model can handle continuous input and output under strict realtime constraints.

\paragraph{Reasoning without sacrificing interaction.} DuplexOmni achieves the best Big Bench Audio score of 77.2\% and remains competitive on Daily-Omni. This suggests that the layered design preserves realtime interaction quality while using the thinking layer to improve audio understanding and complex reasoning.

\begin{table}[ht]
  \centering
  \setlength{\tabcolsep}{4pt}
  \begin{tabular*}{\columnwidth}{@{\extracolsep{\fill}}lcc}
    \toprule
    Model &
    \begin{tabular}[c]{@{}c@{}}DuplexBench\\ToR (\%)\end{tabular} &
    \begin{tabular}[c]{@{}c@{}}Big Bench\\Audio (\%)\end{tabular} \\
    \midrule
    DuplexOmni & 72.6 & 77.2 \\
    Weak Thinking & 72.1 & 50.3 \\
    w/o Thinking & 65.2 & 22.2 \\
    w/o Thinking \& ASR & 67.2 & 18.1 \\
    Thinking Only & -- & 58.9 \\
    \bottomrule
  \end{tabular*}
  \caption{Ablation study on DuplexOmni. Full DuplexBench measures interaction quality, while Big Bench Audio evaluates audio understanding under the streaming setting.}
  \label{tab:ablation}
\end{table}

\subsection{Ablation Study}

Table~\ref{tab:ablation} presents the ablation results of DuplexOmni. \textit{Weak Thinking} replaces the default thinking layer with a weaker model. \textit{w/o Thinking} removes the thinking layer and keeps only the interaction layer. \textit{w/o Thinking \& ASR} further removes speech-to-text output. \textit{Thinking Only} directly uses the thinking layer to process audio input without the DuplexOmni interaction layer. Based on these settings, we obtain three observations.

\paragraph{The full-duplex ability of the interaction layer is independent}. Replacing the thinking layer with a weaker model only changes Full DuplexBench from 72.6\% to 72.1\%. This shows that full-duplex interaction quality is mainly determined by the DuplexOmni model, and is not sensitive to the strength of the thinking layer.

\paragraph{The thinking layer determines the reasoning ceiling.} When the thinking layer is replaced by a weaker model, Big Bench Audio drops from 77.2\% to 50.3\%; removing the thinking layer further reduces it to 22.2\%. Together with the first observation, this shows that the thinking layer is a pluggable component of the interaction layer.

\begin{table}[ht]
  \centering
  \begin{tabular*}{\columnwidth}{@{\extracolsep{\fill}}lccc}
    \toprule
    Length &
    \begin{tabular}[c]{@{}c@{}}Gemini\\Lite\end{tabular} &
    \begin{tabular}[c]{@{}c@{}}Gemini\\Live\end{tabular} &
    \begin{tabular}[c]{@{}c@{}}Duplex\\Omni\end{tabular} \\
    \midrule
    1--5   & 4.9  & 7.8  & 25.1  \\
    6--10  & 5.3  & 4.4  & 15.4  \\
    11--15 & 3.4  & 3.3  & 12.0  \\
    16--20 & 2.9  & 2.2  & 10.6  \\
    21+    & 3.2  & 4.6  & 8.8   \\
    \midrule
    Overall & 3.75 & 4.09 & 11.92 \\
    \bottomrule
  \end{tabular*}
  \caption{ASR error rate by utterance length. Short utterances are substantially harder for DuplexOmni under the full-duplex streaming setting.}
  \label{tab:asr-length-analysis}
\end{table}

\paragraph{The interaction layer improves the effectiveness of the thinking layer.} \textit{Thinking Only} directly processes audio input with the thinking layer and reaches 58.9\% on Big Bench Audio, while the full DuplexOmni reaches 77.2\%. This indicates that the gain does not only come from the external strong model. The interaction layer also organizes and filters the input passed to the thinking layer, making the thinking process more effective.

\subsection{Full-Duplex ASR Analysis}

Table~\ref{tab:asr-length-analysis} shows that DuplexOmni performs worse on short utterances, with WER dropping from 25.1\% for 1--5 words to 8.8\% for 21+ words. This suggests that short, low-context speech fragments in full-duplex interaction remain a key challenge.

\section{Conclusion}
We present DuplexOmni, a real-time multimodal full-duplex interaction method that decouples low-latency interaction from asynchronous thinking and tool use. By combining the DuplexOmni interaction model with a pluggable thinking layer and a Writer-Director data pipeline, DuplexOmni supports continuous listening, seeing, speaking, and reasoning. Experiments show that it achieves strong full-duplex interaction performance while maintaining low response latency.

\section*{Limitations}
DuplexOmni still has several limitations. First, its video capability remains limited because the amount of video-call and visually grounded interaction data is relatively small. Second, its English speech ability is weaker than desired, partly due to the training data being dominated by Chinese speech. We leave stronger video modeling and more balanced multilingual speech training to future work.
\bibliography{custom}

@article{DBLP:journals/corr/abs-2405-19487,
  author       = {Peng Wang and
                  Songshuo Lu and
                  Yaohua Tang and
                  Sijie Yan and
                  Yuanjun Xiong and
                  Wei Xia},
  title        = {A Full-duplex Speech Dialogue Scheme Based On Large Language Models},
  journal      = {CoRR},
  volume       = {abs/2405.19487},
  year         = {2024},
  url          = {https://doi.org/10.48550/arXiv.2405.19487},
  doi          = {10.48550/ARXIV.2405.19487},
  eprinttype   = {arXiv},
  eprint       = {2405.19487},
  bibsource    = {dblp computer science bibliography, https://dblp.org}
}

@article{DBLP:journals/corr/abs-2506-01268,
  author       = {Yudong Lu and
                  Yazhe Niu and
                  Shuai Hu and
                  Haolin Wang},
  title        = {CleanS2S: Single-file Framework for Proactive Speech-to-Speech Interaction},
  journal      = {CoRR},
  volume       = {abs/2506.01268},
  year         = {2025},
  url          = {https://doi.org/10.48550/arXiv.2506.01268},
  doi          = {10.48550/ARXIV.2506.01268},
  eprinttype   = {arXiv},
  eprint       = {2506.01268},
  bibsource    = {dblp computer science bibliography, https://dblp.org}
}

@inproceedings{DBLP:conf/acl/HeldZSLRY25,
  author       = {William Barr Held and
                  Yanzhe Zhang and
                  Weiyan Shi and
                  Minzhi Li and
                  Michael J. Ryan and
                  Diyi Yang},
  editor       = {Wanxiang Che and
                  Joyce Nabende and
                  Ekaterina Shutova and
                  Mohammad Taher Pilehvar},
  title        = {Distilling an End-to-End Voice Assistant Without Instruction Training
                  Data},
  booktitle    = {Proceedings of the 63rd Annual Meeting of the Association for Computational
                  Linguistics (Volume 1: Long Papers), {ACL} 2025, Vienna, Austria,
                  July 27 - August 1, 2025},
  pages        = {7876--7891},
  publisher    = {Association for Computational Linguistics},
  year         = {2025},
  url          = {https://aclanthology.org/2025.acl-long.388/},
  bibsource    = {dblp computer science bibliography, https://dblp.org}
}

@article{DBLP:journals/corr/abs-2509-06502,
  author       = {Junjie Chen and
                  Yao Hu and
                  Junjie Li and
                  Kangyue Li and
                  Kun Liu and
                  Wenpeng Li and
                  Xu Li and
                  Ziyuan Li and
                  Feiyu Shen and
                  Xu Tang and
                  Manzhen Wei and
                  Yichen Wu and
                  Fenglong Xie and
                  Kaituo Xu and
                  Kun Xie},
  title        = {FireRedChat: {A} Pluggable, Full-Duplex Voice Interaction System with
                  Cascaded and Semi-Cascaded Implementations},
  journal      = {CoRR},
  volume       = {abs/2509.06502},
  year         = {2025},
  url          = {https://doi.org/10.48550/arXiv.2509.06502},
  doi          = {10.48550/ARXIV.2509.06502},
  eprinttype   = {arXiv},
  eprint       = {2509.06502},
  bibsource    = {dblp computer science bibliography, https://dblp.org}
}

@article{DBLP:journals/corr/abs-2512-20156,
  author       = {Tongyi Fun Team and
                  Qian Chen and
                  Luyao Cheng and
                  Chong Deng and
                  Xiangang Li and
                  Jiaqing Liu and
                  Chao{-}Hong Tan and
                  Wen Wang and
                  Junhao Xu and
                  Jieping Ye and
                  Qinglin Zhang and
                  Qiquan Zhang and
                  Jingren Zhou},
  title        = {Fun-Audio-Chat Technical Report},
  journal      = {CoRR},
  volume       = {abs/2512.20156},
  year         = {2025},
  url          = {https://doi.org/10.48550/arXiv.2512.20156},
  doi          = {10.48550/ARXIV.2512.20156},
  eprinttype   = {arXiv},
  eprint       = {2512.20156},
  bibsource    = {dblp computer science bibliography, https://dblp.org}
}

@inproceedings{DBLP:conf/aaai/MaSDC0WW025,
  author       = {Ziyang Ma and
                  Yakun Song and
                  Chenpeng Du and
                  Jian Cong and
                  Zhuo Chen and
                  Yuping Wang and
                  Yuxuan Wang and
                  Xie Chen},
  editor       = {Toby Walsh and
                  Julie Shah and
                  Zico Kolter},
  title        = {Language Model Can Listen While Speaking},
  booktitle    = {Thirty-Ninth {AAAI} Conference on Artificial Intelligence, Thirty-Seventh
                  Conference on Innovative Applications of Artificial Intelligence,
                  Fifteenth Symposium on Educational Advances in Artificial Intelligence,
                  {AAAI} 2025, Philadelphia, PA, USA, February 25 - March 4, 2025},
  pages        = {24831--24839},
  publisher    = {{AAAI} Press},
  year         = {2025},
  url          = {https://doi.org/10.1609/aaai.v39i23.34665},
  doi          = {10.1609/AAAI.V39I23.34665},
  bibsource    = {dblp computer science bibliography, https://dblp.org}
}

@article{DBLP:journals/corr/abs-2508-15827,
  author       = {Zhifei Xie and
                  Ziyang Ma and
                  Zihang Liu and
                  Kaiyu Pang and
                  Hongyu Li and
                  Jialin Zhang and
                  Yue Liao and
                  Deheng Ye and
                  Chunyan Miao and
                  Shuicheng Yan},
  title        = {Mini-Omni-Reasoner: Token-Level Thinking-in-Speaking in Large Speech
                  Models},
  journal      = {CoRR},
  volume       = {abs/2508.15827},
  year         = {2025},
  url          = {https://doi.org/10.48550/arXiv.2508.15827},
  doi          = {10.48550/ARXIV.2508.15827},
  eprinttype   = {arXiv},
  eprint       = {2508.15827},
  bibsource    = {dblp computer science bibliography, https://dblp.org}
}

@inproceedings{DBLP:conf/acl/ZhangCDCWZLYTDZ25,
  author       = {Qinglin Zhang and
                  Luyao Cheng and
                  Chong Deng and
                  Qian Chen and
                  Wen Wang and
                  Siqi Zheng and
                  Jiaqing Liu and
                  Hai Yu and
                  Chao{-}Hong Tan and
                  Zhihao Du and
                  Shiliang Zhang},
  editor       = {Wanxiang Che and
                  Joyce Nabende and
                  Ekaterina Shutova and
                  Mohammad Taher Pilehvar},
  title        = {OmniFlatten: An End-to-end {GPT} Model for Seamless Voice Conversation},
  booktitle    = {Proceedings of the 63rd Annual Meeting of the Association for Computational
                  Linguistics (Volume 1: Long Papers), {ACL} 2025, Vienna, Austria,
                  July 27 - August 1, 2025},
  pages        = {14570--14580},
  publisher    = {Association for Computational Linguistics},
  year         = {2025},
  url          = {https://aclanthology.org/2025.acl-long.709/},
  bibsource    = {dblp computer science bibliography, https://dblp.org}
}

@inproceedings{DBLP:conf/acl/Chen0YLLXN00L0025,
  author       = {Wenxi Chen and
                  Ziyang Ma and
                  Ruiqi Yan and
                  Yuzhe Liang and
                  Xiquan Li and
                  Ruiyang Xu and
                  Zhikang Niu and
                  Yanqiao Zhu and
                  Yifan Yang and
                  Zhanxun Liu and
                  Kai Yu and
                  Yuxuan Hu and
                  Jinyu Li and
                  Yan Lu and
                  Shujie Liu and
                  Xie Chen},
  editor       = {Wanxiang Che and
                  Joyce Nabende and
                  Ekaterina Shutova and
                  Mohammad Taher Pilehvar},
  title        = {SLAM-Omni: Timbre-Controllable Voice Interaction System with Single-Stage
                  Training},
  booktitle    = {Findings of the Association for Computational Linguistics, {ACL} 2025,
                  Vienna, Austria, July 27 - August 1, 2025},
  series       = {Findings of {ACL}},
  pages        = {2262--2282},
  publisher    = {Association for Computational Linguistics},
  year         = {2025},
  url          = {https://aclanthology.org/2025.findings-acl.115/},
  bibsource    = {dblp computer science bibliography, https://dblp.org}
}

@article{DBLP:journals/corr/abs-2410-00037,
  author       = {Alexandre D{\'{e}}fossez and
                  Laurent Mazar{\'{e}} and
                  Manu Orsini and
                  Am{\'{e}}lie Royer and
                  Patrick P{\'{e}}rez and
                  Herv{\'{e}} J{\'{e}}gou and
                  Edouard Grave and
                  Neil Zeghidour},
  title        = {Moshi: a speech-text foundation model for real-time dialogue},
  journal      = {CoRR},
  volume       = {abs/2410.00037},
  year         = {2024},
  url          = {https://doi.org/10.48550/arXiv.2410.00037},
  doi          = {10.48550/ARXIV.2410.00037},
  eprinttype   = {arXiv},
  eprint       = {2410.00037},
  bibsource    = {dblp computer science bibliography, https://dblp.org}
}

@article{DBLP:journals/corr/abs-2604-15804,
  author       = {Qwen Team},
  title        = {Qwen3.5-Omni Technical Report},
  journal      = {CoRR},
  volume       = {abs/2604.15804},
  year         = {2026},
  url          = {https://doi.org/10.48550/arXiv.2604.15804},
  doi          = {10.48550/ARXIV.2604.15804},
  eprinttype   = {arXiv},
  eprint       = {2604.15804},
  bibsource    = {dblp computer science bibliography, https://dblp.org}
}

@article{DBLP:journals/corr/abs-2509-17765,
  author       = {Qwen Team},
  title        = {Qwen3-Omni Technical Report},
  journal      = {CoRR},
  volume       = {abs/2509.17765},
  year         = {2025},
  url          = {https://doi.org/10.48550/arXiv.2509.17765},
  doi          = {10.48550/ARXIV.2509.17765},
  eprinttype   = {arXiv},
  eprint       = {2509.17765},
  bibsource    = {dblp computer science bibliography, https://dblp.org}
}

@inproceedings{DBLP:conf/asru/LinLLWALL25,
  author       = {Guan{-}Ting Lin and
                  Jiachen Lian and
                  Tingle Li and
                  Qirui Wang and
                  Gopala Anumanchipalli and
                  Alexander H. Liu and
                  Hung{-}Yi Lee},
  title        = {Full-Duplex-Bench: {A} Benchmark to Evaluate Full-Duplex Spoken Dialogue
                  Models on Turn-taking Capabilities},
  booktitle    = {{IEEE} Automatic Speech Recognition and Understanding Workshop, {ASRU}
                  2025, Honolulu, HI, USA, December 6-10, 2025},
  pages        = {1--8},
  publisher    = {{IEEE}},
  year         = {2025},
  url          = {https://doi.org/10.1109/ASRU65441.2025.11433838},
  doi          = {10.1109/ASRU65441.2025.11433838},
  bibsource    = {dblp computer science bibliography, https://dblp.org}
}

@article{srivastava2022beyond,
  title={Beyond the Imitation Game: Quantifying and extrapolating the capabilities of language models},
  author={Srivastava, Aarohi and Rastogi, Abhinav and Rao, Abhishek and Shoeb, Abu Awal Md and Abid, Abubakar and Fisch, Adam and Brown, Adam R and Santoro, Adam and Gupta, Aditya and Garriga-Alonso, Adri{\`a} and others},
  journal={arXiv preprint arXiv:2206.04615},
  year={2022}
}

@article{suzgun2022challenging,
  title={Challenging BIG-Bench Tasks and Whether Chain-of-Thought Can Solve Them},
  author={Suzgun, Mirac and Scales, Nathan and Sch{\"a}rli, Nathanael and Gehrmann, Sebastian and Tay, Yi and Chung, Hyung Won and Chowdhery, Aakanksha and Le, Quoc V and Chi, Ed H and Zhou, Denny and Wei, Jason},
  journal={arXiv preprint arXiv:2210.09261},
  year={2022}
}

@article{DBLP:journals/corr/abs-2505-17862,
  author       = {Ziwei Zhou and
                  Rui Wang and
                  Zuxuan Wu},
  title        = {Daily-Omni: Towards Audio-Visual Reasoning with Temporal Alignment
                  across Modalities},
  journal      = {CoRR},
  volume       = {abs/2505.17862},
  year         = {2025},
  url          = {https://doi.org/10.48550/arXiv.2505.17862},
  doi          = {10.48550/ARXIV.2505.17862},
  eprinttype   = {arXiv},
  eprint       = {2505.17862},
  bibsource    = {dblp computer science bibliography, https://dblp.org}
}

@inproceedings{DBLP:conf/icassp/PanayotovCPK15,
  author       = {Vassil Panayotov and
                  Guoguo Chen and
                  Daniel Povey and
                  Sanjeev Khudanpur},
  title        = {Librispeech: An {ASR} corpus based on public domain audio books},
  booktitle    = {2015 {IEEE} International Conference on Acoustics, Speech and Signal
                  Processing, {ICASSP} 2015, South Brisbane, Queensland, Australia,
                  April 19-24, 2015},
  pages        = {5206--5210},
  publisher    = {{IEEE}},
  year         = {2015},
  url          = {https://doi.org/10.1109/ICASSP.2015.7178964},
  doi          = {10.1109/ICASSP.2015.7178964},
  bibsource    = {dblp computer science bibliography, https://dblp.org}
}

@article{DBLP:journals/corr/abs-2410-21276,
  author       = {OpenAI},
  title        = {GPT-4o System Card},
  journal      = {CoRR},
  volume       = {abs/2410.21276},
  year         = {2024},
  url          = {https://doi.org/10.48550/arXiv.2410.21276},
  doi          = {10.48550/ARXIV.2410.21276},
  eprinttype   = {arXiv},
  eprint       = {2410.21276},
  bibsource    = {dblp computer science bibliography, https://dblp.org}
}

@article{DBLP:journals/corr/abs-2312-11805,
  author       = {Gemini Team},
  title        = {Gemini: {A} Family of Highly Capable Multimodal Models},
  journal      = {CoRR},
  volume       = {abs/2312.11805},
  year         = {2023},
  url          = {https://doi.org/10.48550/arXiv.2312.11805},
  doi          = {10.48550/ARXIV.2312.11805},
  eprinttype   = {arXiv},
  eprint       = {2312.11805},
  bibsource    = {dblp computer science bibliography, https://dblp.org}
}

@article{DBLP:journals/corr/abs-2403-05530,
  author       = {Machel Reid and
                  Nikolay Savinov and
                  Denis Teplyashin and
                  Dmitry Lepikhin and
                  Timothy P. Lillicrap and
                  Jean{-}Baptiste Alayrac and
                  Radu Soricut and
                  Angeliki Lazaridou and
                  Orhan Firat and
                  Julian Schrittwieser and
                  Ioannis Antonoglou and
                  Rohan Anil and
                  Sebastian Borgeaud and
                  Andrew M. Dai and
                  Katie Millican and
                  Ethan Dyer and
                  Mia Glaese and
                  Thibault Sottiaux and
                  Benjamin Lee and
                  Fabio Viola and
                  Malcolm Reynolds and
                  Yuanzhong Xu and
                  James Molloy and
                  Jilin Chen and
                  Michael Isard and
                  Paul Barham and
                  Tom Hennigan and
                  Ross McIlroy and
                  Melvin Johnson and
                  Johan Schalkwyk and
                  Eli Collins and
                  Eliza Rutherford and
                  Erica Moreira and
                  Kareem Ayoub and
                  Megha Goel and
                  Clemens Meyer and
                  Gregory Thornton and
                  Zhen Yang and
                  Henryk Michalewski and
                  Zaheer Abbas and
                  Nathan Schucher and
                  Ankesh Anand and
                  Richard Ives and
                  James Keeling and
                  Karel Lenc and
                  Salem Haykal and
                  Siamak Shakeri and
                  Pranav Shyam and
                  Aakanksha Chowdhery and
                  Roman Ring and
                  Stephen Spencer and
                  Eren Sezener and
                  et al.},
  title        = {Gemini 1.5: Unlocking multimodal understanding across millions of
                  tokens of context},
  journal      = {CoRR},
  volume       = {abs/2403.05530},
  year         = {2024},
  url          = {https://doi.org/10.48550/arXiv.2403.05530},
  doi          = {10.48550/ARXIV.2403.05530},
  eprinttype   = {arXiv},
  eprint       = {2403.05530},
  bibsource    = {dblp computer science bibliography, https://dblp.org}
}

@article{DBLP:journals/corr/abs-2503-20215,
  author       = {Jin Xu and
                  Zhifang Guo and
                  Jinzheng He and
                  Hangrui Hu and
                  Ting He and
                  Shuai Bai and
                  Keqin Chen and
                  Jialin Wang and
                  Yang Fan and
                  Kai Dang and
                  Bin Zhang and
                  Xiong Wang and
                  Yunfei Chu and
                  Junyang Lin},
  title        = {Qwen2.5-Omni Technical Report},
  journal      = {CoRR},
  volume       = {abs/2503.20215},
  year         = {2025},
  url          = {https://doi.org/10.48550/arXiv.2503.20215},
  doi          = {10.48550/ARXIV.2503.20215},
  eprinttype   = {arXiv},
  eprint       = {2503.20215},
  bibsource    = {dblp computer science bibliography, https://dblp.org}
}

@article{yao2024minicpm,
  title={MiniCPM-V: A GPT-4V Level MLLM on Your Phone},
  author={Yao, Yuan and Yu, Tianyu and Zhang, Ao and Wang, Chongyi and Cui, Junbo and Zhu, Hongji and Cai, Tianchi and Li, Haoyu and Zhao, Weilin and He, Zhihui and others},
  journal={arXiv preprint arXiv:2408.01800},
  year={2024}
}

@article{DBLP:journals/corr/abs-2501-15368,
  author       = {Baichuan Inc.},
  title        = {Baichuan-Omni-1.5 Technical Report},
  journal      = {CoRR},
  volume       = {abs/2501.15368},
  year         = {2025},
  url          = {https://doi.org/10.48550/arXiv.2501.15368},
  doi          = {10.48550/ARXIV.2501.15368},
  eprinttype   = {arXiv},
  eprint       = {2501.15368},
  bibsource    = {dblp computer science bibliography, https://dblp.org}
}

@article{DBLP:journals/corr/abs-2504-18425,
  author       = {KimiTeam and
                  Ding Ding and
                  Zeqian Ju and
                  Yichong Leng and
                  Songxiang Liu and
                  Tong Liu and
                  Zeyu Shang and
                  Kai Shen and
                  Wei Song and
                  Xu Tan and
                  Heyi Tang and
                  Zhengtao Wang and
                  Chu Wei and
                  Yifei Xin and
                  Xinran Xu and
                  Jianwei Yu and
                  Yutao Zhang and
                  Xinyu Zhou and
                  Y. Charles and
                  Jun Chen and
                  Yanru Chen and
                  Yulun Du and
                  Weiran He and
                  Zhenxing Hu and
                  Guokun Lai and
                  Qingcheng Li and
                  Yangyang Liu and
                  Weidong Sun and
                  Jianzhou Wang and
                  Yuzhi Wang and
                  Yuefeng Wu and
                  Yuxin Wu and
                  Dongchao Yang and
                  Hao Yang and
                  Ying Yang and
                  Zhilin Yang and
                  Aoxiong Yin and
                  Ruibin Yuan and
                  Yutong Zhang and
                  Zaida Zhou},
  title        = {Kimi-Audio Technical Report},
  journal      = {CoRR},
  volume       = {abs/2504.18425},
  year         = {2025},
  url          = {https://doi.org/10.48550/arXiv.2504.18425},
  doi          = {10.48550/ARXIV.2504.18425},
  eprinttype   = {arXiv},
  eprint       = {2504.18425},
  bibsource    = {dblp computer science bibliography, https://dblp.org}
}

@article{DBLP:journals/corr/abs-2407-10759,
  author       = {Yunfei Chu and
                  Jin Xu and
                  Qian Yang and
                  Haojie Wei and
                  Xipin Wei and
                  Zhifang Guo and
                  Yichong Leng and
                  Yuanjun Lv and
                  Jinzheng He and
                  Junyang Lin and
                  Chang Zhou and
                  Jingren Zhou},
  title        = {Qwen2-Audio Technical Report},
  journal      = {CoRR},
  volume       = {abs/2407.10759},
  year         = {2024},
  url          = {https://doi.org/10.48550/arXiv.2407.10759},
  doi          = {10.48550/ARXIV.2407.10759},
  eprinttype   = {arXiv},
  eprint       = {2407.10759},
  bibsource    = {dblp computer science bibliography, https://dblp.org}
}

@inproceedings{DBLP:conf/emnlp/ZhangLZZWZQ23,
  author       = {Dong Zhang and
                  Shimin Li and
                  Xin Zhang and
                  Jun Zhan and
                  Pengyu Wang and
                  Yaqian Zhou and
                  Xipeng Qiu},
  editor       = {Houda Bouamor and
                  Juan Pino and
                  Kalika Bali},
  title        = {SpeechGPT: Empowering Large Language Models with Intrinsic Cross-Modal
                  Conversational Abilities},
  booktitle    = {Findings of the Association for Computational Linguistics: {EMNLP}
                  2023, Singapore, December 6-10, 2023},
  series       = {Findings of {ACL}},
  pages        = {15757--15773},
  publisher    = {Association for Computational Linguistics},
  year         = {2023},
  url          = {https://doi.org/10.18653/v1/2023.findings-emnlp.1055},
  doi          = {10.18653/V1/2023.FINDINGS-EMNLP.1055},
  bibsource    = {dblp computer science bibliography, https://dblp.org}
}

@article{DBLP:journals/corr/abs-2603-09180,
  author       = {Jianing Yang and
                  Yusuke Fujita and
                  Yui Sudo},
  title        = {DuplexCascade: Full-Duplex Speech-to-Speech Dialogue with VAD-Free
                  Cascaded {ASR-LLM-TTS} Pipeline and Micro-Turn Optimization},
  journal      = {CoRR},
  volume       = {abs/2603.09180},
  year         = {2026},
  url          = {https://doi.org/10.48550/arXiv.2603.09180},
  doi          = {10.48550/ARXIV.2603.09180},
  eprinttype   = {arXiv},
  eprint       = {2603.09180},
  bibsource    = {dblp computer science bibliography, https://dblp.org}
}

@misc{nguyen2022generativespokendialoguelanguage,
      title={Generative Spoken Dialogue Language Modeling}, 
      author={Tu Anh Nguyen and Eugene Kharitonov and Jade Copet and Yossi Adi and Wei-Ning Hsu and Ali Elkahky and Paden Tomasello and Robin Algayres and Benoit Sagot and Abdelrahman Mohamed and Emmanuel Dupoux},
      year={2022},
      eprint={2203.16502},
      archivePrefix={arXiv},
      primaryClass={cs.CL},
      url={https://arxiv.org/abs/2203.16502}, 
}

@article{DBLP:journals/corr/abs-2505-17060,
  author       = {Wenyi Yu and
                  Siyin Wang and
                  Xiaoyu Yang and
                  Xianzhao Chen and
                  Xiaohai Tian and
                  Jun Zhang and
                  Guangzhi Sun and
                  Lu Lu and
                  Yuxuan Wang and
                  Chao Zhang},
  title        = {SALMONN-omni: {A} Standalone Speech {LLM} without Codec Injection
                  for Full-duplex Conversation},
  journal      = {CoRR},
  volume       = {abs/2505.17060},
  year         = {2025},
  url          = {https://doi.org/10.48550/arXiv.2505.17060},
  doi          = {10.48550/ARXIV.2505.17060},
  eprinttype   = {arXiv},
  eprint       = {2505.17060},
  bibsource    = {dblp computer science bibliography, https://dblp.org}
}

@article{DBLP:journals/corr/abs-2512-23808,
  author       = {LLM{-}Core Xiaomi},
  title        = {MiMo-Audio: Audio Language Models are Few-Shot Learners},
  journal      = {CoRR},
  volume       = {abs/2512.23808},
  year         = {2025},
  url          = {https://doi.org/10.48550/arXiv.2512.23808},
  doi          = {10.48550/ARXIV.2512.23808},
  eprinttype   = {arXiv},
  eprint       = {2512.23808},
  bibsource    = {dblp computer science bibliography, https://dblp.org}
}

@article{DBLP:journals/corr/abs-2601-15621,
  author       = {Qwen Team},
  title        = {Qwen3-TTS Technical Report},
  journal      = {CoRR},
  volume       = {abs/2601.15621},
  year         = {2026},
  url          = {https://doi.org/10.48550/arXiv.2601.15621},
  doi          = {10.48550/ARXIV.2601.15621},
  eprinttype   = {arXiv},
  eprint       = {2601.15621},
  bibsource    = {dblp computer science bibliography, https://dblp.org}
}

@article{DBLP:journals/corr/abs-2505-09388,
  author       = {Qwen Team},
  title        = {Qwen3 Technical Report},
  journal      = {CoRR},
  volume       = {abs/2505.09388},
  year         = {2025},
  url          = {https://doi.org/10.48550/arXiv.2505.09388},
  doi          = {10.48550/ARXIV.2505.09388},
  eprinttype   = {arXiv},
  eprint       = {2505.09388},
  bibsource    = {dblp computer science bibliography, https://dblp.org}
}

@article{ding2023enhancing,
  title={Enhancing Chat Language Models by Scaling High-quality Instructional Conversations},
  author={Ding, Ning and Chen, Yulin and Xu, Bokai and Qin, Yujia and Zheng, Zhi and Hu, Shengding and Liu, Zhiyuan and Sun, Maosong and Zhou, Bowen},
  journal={arXiv preprint arXiv:2305.14233},
  year={2023}
}

@misc{zhao2024swiftascalablelightweightinfrastructure,
      title={SWIFT:A Scalable lightWeight Infrastructure for Fine-Tuning},
      author={Yuze Zhao and Jintao Huang and Jinghan Hu and Xingjun Wang and Yunlin Mao and Daoze Zhang and Zeyinzi Jiang and Zhikai Wu and Baole Ai and Ang Wang and Wenmeng Zhou and Yingda Chen},
      year={2024},
      eprint={2408.05517},
      archivePrefix={arXiv},
      primaryClass={cs.CL},
      url={https://arxiv.org/abs/2408.05517},
}

@misc{zhao2024wildchat1mchatgptinteraction,
      title={WildChat: 1M ChatGPT Interaction Logs in the Wild}, 
      author={Wenting Zhao and Xiang Ren and Jack Hessel and Claire Cardie and Yejin Choi and Yuntian Deng},
      year={2024},
      eprint={2405.01470},
      archivePrefix={arXiv},
      primaryClass={cs.CL},
      url={https://arxiv.org/abs/2405.01470}, 
}

@article{DBLP:journals/corr/abs-2304-07854,
  author       = {Yunjie Ji and
                  Yan Gong and
                  Yong Deng and
                  Yiping Peng and
                  Qiang Niu and
                  Baochang Ma and
                  Xiangang Li},
  title        = {Towards Better Instruction Following Language Models for Chinese:
                  Investigating the Impact of Training Data and Evaluation},
  journal      = {CoRR},
  volume       = {abs/2304.07854},
  year         = {2023},
  url          = {https://doi.org/10.48550/arXiv.2304.07854},
  doi          = {10.48550/ARXIV.2304.07854},
  eprinttype   = {arXiv},
  eprint       = {2304.07854},
  bibsource    = {dblp computer science bibliography, https://dblp.org}
}

@misc{zhang2023chineseopeninstructiongeneralist,
      title={Chinese Open Instruction Generalist: A Preliminary Release}, 
      author={Ge Zhang and Yemin Shi and Ruibo Liu and Ruibin Yuan and Yizhi Li and Siwei Dong and Yu Shu and Zhaoqun Li and Zekun Wang and Chenghua Lin and Wenhao Huang and Jie Fu},
      year={2023},
      eprint={2304.07987},
      archivePrefix={arXiv},
      primaryClass={cs.CL},
      url={https://arxiv.org/abs/2304.07987}, 
}

@misc{no_robots,
  author = {Nazneen Rajani and Lewis Tunstall and Edward Beeching and Nathan Lambert and Alexander M. Rush and Thomas Wolf},
  title = {No Robots},
  year = {2023},
  publisher = {Hugging Face},
  journal = {Hugging Face repository},
  howpublished = {\url{https://huggingface.co/datasets/HuggingFaceH4/no_robots}}
}

@misc{kpf2023openassistantconversationsdemocratizing,
      title={OpenAssistant Conversations -- Democratizing Large Language Model Alignment}, 
      author={Andreas Köpf and Yannic Kilcher and Dimitri von Rütte and Sotiris Anagnostidis and Zhi-Rui Tam and Keith Stevens and Abdullah Barhoum and Nguyen Minh Duc and Oliver Stanley and Richárd Nagyfi and Shahul ES and Sameer Suri and David Glushkov and Arnav Dantuluri and Andrew Maguire and Christoph Schuhmann and Huu Nguyen and Alexander Mattick},
      year={2023},
      eprint={2304.07327},
      archivePrefix={arXiv},
      primaryClass={cs.CL},
      url={https://arxiv.org/abs/2304.07327}, 
}
\appendix
\section{Full Control Token Set}
\label{sec:appendix-tokens}

Table~\ref{tab:control-tokens} lists the complete set of Director annotation tokens inserted during the Writer-Director pipeline. These tokens encode the temporal relationship between user speech, assistant speech, and background thinking in each training sample.

\begin{table*}[ht]
\centering
\small
\setlength{\tabcolsep}{5pt}
\renewcommand{\arraystretch}{1.2}
\begin{tabular}{p{1.9cm} p{1.4cm} p{6.0cm} p{5.6cm}}
\toprule
\textbf{Token} & \textbf{Turn} & \textbf{Semantics} & \textbf{Placement} \\
\midrule
\texttt{[THINK]}           & User      & Triggers S2 to begin background reasoning.                           & End of the user utterance that requires external reasoning. \\
\texttt{\textless{}\ldots{}\textgreater{}}       & Assistant & Injects a complete S2 result fragment before the corresponding fact. & Before the relevant assistant sentence; never as the first token after a user turn. \\
\texttt{\string^}          & Assistant & Marks the character offset where the next speaker begins.            & Inside the assistant text at the overlap onset; agent cannot trigger user or itself. \\
\texttt{[CUT]}             & Assistant & Marks where assistant audio stops; following text is ghost text.     & 2--4 characters after \texttt{\string^}; omitted for pure backchannel overlap. \\
\texttt{[WAIT]}            & User      & Suspends and resets the ongoing S2 reasoning process.                & End of the interrupting user turn; a new \texttt{[THINK]} is required to resume. \\
\texttt{[PEND\textit{N}S]} & User      & Encodes $N$ seconds of shared silence ($N\!=\!1$--$35$).            & Start of the user turn that resumes after the silence; forbidden in assistant turns. \\
\bottomrule
\end{tabular}
\caption{Complete set of Director annotation tokens. Ghost text (after \texttt{[CUT]}) preserves what the assistant intended but did not utter.}
\label{tab:control-tokens}
\end{table*}

\begin{table*}[ht]
\small
\centering
\renewcommand{\arraystretch}{1.2}
\begin{tabular}{lc}
\toprule
\textbf{Interaction Pattern} & \textbf{Occurrence (\%)} \\
\midrule
Delayed reasoning (\texttt{[THINK]} + S2 fragments) & 94.3 \\
Shared silence (\texttt{[PEND\textit{N}S]})          & 68.2 \\
Assistant-initiated turn                              & 50.0 \\
Overlapping speech (\texttt{\string^} + \texttt{[CUT]}) & 49.8 \\
Interruption with reset (\texttt{[CUT]} + \texttt{[WAIT]}) & 41.9 \\
Pure backchannel (\texttt{\string^} without \texttt{[CUT]}) &  3.1 \\
\midrule
Samples containing $\geq$2 patterns                  & 90.7 \\
\bottomrule
\end{tabular}
\caption{Proportion of training samples containing each interaction pattern. Categories are non-exclusive; 90.7\% of samples exhibit two or more patterns simultaneously.}
\label{tab:token-distribution}
\end{table*}

\begin{table}[ht]
\centering
\renewcommand{\arraystretch}{1.2}
\begin{tabular}{lc}
\toprule
\textbf{Scenario Type} & \textbf{Ratio (\%)} \\
\midrule
Step-by-step guidance    & 32.1 \\
Standard turn-taking     & 31.3 \\
Condition revision       & 18.4 \\
Troubleshooting guidance & 13.9 \\
Extended listening       &  4.4 \\
\bottomrule
\end{tabular}
\caption{Distribution of primary interaction types in DuplexOmni scenario seeds. The distribution is intentionally diversified to expose the model to different temporal structures and improve robustness under continuous real-time interaction.}
\label{tab:scenario-distribution}
\end{table}
\begin{table}[ht]
\centering
\renewcommand{\arraystretch}{1.2}
\begin{tabular}{lc}
\toprule
\textbf{Statistic} & \textbf{Value} \\
\midrule
Total scenario seeds              & $\sim$620K \\
Total raw conversations           & $\sim$3.02M \\
Average turns per sample          & 10.4 \\
Turn range                        & 4--33 \\
S2 participation: high            & 23.6\% \\
S2 participation: medium          & 46.0\% \\
S2 participation: low             & 24.9\% \\
S2 participation: off             &  5.5\% \\
\bottomrule
\end{tabular}
\caption{Key statistics of the DuplexOmni training corpus after Writer-Director annotation.}
\label{tab:data-stats}
\end{table}
The six tokens capture five distinct temporal phenomena in full-duplex interaction: (1)~\textbf{thinking onset} (\texttt{[THINK]}), triggering asynchronous background reasoning at a precise moment in the user turn; (2)~\textbf{thinking delivery} (\texttt{\textless{}\ldots{}\textgreater{}}), injecting reasoning results into the assistant stream with explicit causal ordering; (3)~\textbf{speech overlap onset} (\texttt{\string^}), recording the exact character-level time point at which a second speaker begins; (4)~\textbf{speech cutoff} (\texttt{[CUT]}), recording where the first speaker stops while preserving ghost text for loss computation; (5)~\textbf{thinking suspension} (\texttt{[WAIT]}), resetting the reasoning state when user intent changes; and (6)~\textbf{shared silence} (\texttt{[PEND\textit{N}S]}), encoding mutual pauses with second-level granularity.

These tokens are stripped from the text before TTS synthesis and force-alignment; the resulting character-level timestamps are used to construct the time-sliced training samples consumed by the DuplexOmni model.

\section{Examples of Writer-Director Outputs}
\label{sec:appendix-examples}

The following four examples are drawn verbatim from actual Writer-Director pipeline outputs.
Each example is a complete annotated dialogue produced by the Director stage for one scenario seed.
Control tokens appear inline in \texttt{monospace}; \texttt{\textless{}\ldots{}\textgreater{}} carries the S2 reasoning fragment that precedes each factual assistant utterance.

\paragraph{Example 1: Interruption with condition change.}
The user requests a story (\texttt{[THINK]}) but interrupts the assistant before it can finish asking a clarifying question.
\texttt{\string^} marks the moment the user begins speaking during the assistant's turn; \texttt{[CUT]} marks where the assistant's audio actually stops (ghost text follows).
\texttt{[WAIT]} resets the ongoing S2 reasoning; a new \texttt{[THINK]} later restarts it when the user adds further detail.
The conversation also contains a one-second shared silence (\texttt{[PEND1S]}) while the user thinks.

\begin{quote}
\small
\textbf{User:} Um, I wanna hear a story. \texttt{[THINK]}

\textbf{Assistant:} No problem. What kind of \texttt{\string^}topic\texttt{[CUT]} did you have in mind?

\textbf{User:} Warm and caring---family helping strangers. No need to ask so many questions. \texttt{[WAIT]}

\textbf{Assistant:} I need to confirm the tone, or the output might be off. \texttt{[THINK]}

\textbf{User:} Fine, just confirm already.

\textbf{Assistant:} OK, wait a sec.

\textbf{Assistant:} \texttt{\textless{}The story's tone is warm and healing, initially set on a street.\textgreater{}}Warm and comforting tone. Start on a street.

\textbf{User:} Wait, the street's too cold. Let's change to a restaurant.

\textbf{Assistant:} Got it. Switching to restaurant scene.

\textbf{User:} \texttt{[PEND1S]} He needs to suffer more, or it won't feel real. \texttt{[THINK]}

\textbf{Assistant:} Let me check the previous settings to avoid conflicts.

\textbf{Assistant:} \texttt{\textless{}The man is unemployed and down on his luck; his family treats him to a meal at a restaurant.\textgreater{}}Fixed. A man, jobless and down on his luck, is invited to dinner by his family. Is that okay?
\end{quote}

\paragraph{Example 2: Backchannel without speech cutoff.}
The user acknowledges mid-stream while the assistant is describing a product feature.
\texttt{\string^} records the overlap onset, but no \texttt{[CUT]} follows because the assistant's speech continues uninterrupted---this is a non-disruptive backchannel.
Multiple \texttt{\textless{}\ldots{}\textgreater{}} fragments arrive in sequence, pacing the assistant's factual delivery.

\begin{quote}
\small
\textbf{User:} Hi, if the site doesn't work on mobile, will SEO be affected? \texttt{[THINK]}

\textbf{Assistant:} I'll check that.

\textbf{Assistant:} \texttt{\textless{}Mobile-friendliness is a key ranking factor; non-mobile-friendly sites will have reduced search weight.\textgreater{}}It really matters. Most traffic now comes from phones, so search engines favor mobile-friendly sites.

\textbf{User:} Can you help me check my website?

\textbf{Assistant:} Sure, send me the URL and I'll start the check.

\textbf{User:} It's www.example.com. \texttt{[THINK]}

\textbf{Assistant:} Got it, connecting to the server for\texttt{\string^} verification.

\textbf{User:} Yeah.

\textbf{Assistant:} May take a moment to load the page snapshot.

\textbf{Assistant:} \texttt{\textless{}The site is not fetching real-time data; likely a network or server response issue.\textgreater{}}Sorry, can't fetch real-time data from this site---probably a network issue. Let me try again\ldots{} Still can't connect. Server response seems to be the problem.

\textbf{User:} Why won't it connect? Just tell me how to fix it.

\textbf{Assistant:} Got it. Use responsive design and compress images for faster mobile loading.
\end{quote}

\paragraph{Example 3: Delayed thinking feedback with shared silence.}
The user shifts the question mid-task (\texttt{[WAIT][THINK]}) and twice falls silent before continuing (\texttt{[PEND2S]}, \texttt{[PEND3S]}).
S2 returns two short fragments in sequence, reflecting the ``small-chunk'' delivery mode.
When the user corrects a factual error, a new \texttt{[THINK]} re-queries S2, and the assistant acknowledges the mistake.

\begin{quote}
\small
\textbf{User:} Hey, you busy? Got a quick question---heard of a book called \textit{Young Soul Rebels}? \texttt{[THINK]}

\textbf{Assistant:} I'm free. About this book, I'll answer in two parts: main content first, then social context. Just a moment while I check the database.

\textbf{User:} \texttt{[PEND2S]} Wait, skip the summary. I just want to know why Northern Soul rose---especially the drug culture part. Is that really detailed in the book? \texttt{[THINK]}

\textbf{Assistant:} Got it. I'll recheck the key sections, focusing on drug culture. It'll take a moment.

\textbf{User:} \texttt{[PEND3S]} Fine, hurry up. Nothing vague this time.

\textbf{Assistant:} Found it.

\textbf{Assistant:} \texttt{\textless{}The book mentions that dancers in the 1970s used amphetamines to dance all night, prompting police raids on nightclubs.\textgreater{}}The book mentions dancers used amphetamines to dance all night in the `70s, leading police to raid clubs.

\textbf{Assistant:} \texttt{\textless{}The Twisted Wheel club in Manchester was raided.\textgreater{}}For example, Manchester's Twisted Wheel Club got raided.

\textbf{User:} Twisted Wheel? I thought the Wigan Casino was more famous. Did you mix them up? \texttt{[THINK]}

\textbf{Assistant:} Sorry, I was going on an initial search. Since you mentioned Wigan Casino, let me double-check.

\textbf{Assistant:} \texttt{\textless{}The Wigan Casino was frequently raided due to overcrowding and drug use.\textgreater{}}You're right---the Wigan Casino has been raided multiple times over crowd and drug issues. The earlier info was inaccurate.
\end{quote}

\paragraph{Example 4: Assistant-initiated outbound dialogue.}
The assistant opens the conversation proactively.
The user's first turn includes a two-second silence (\texttt{[PEND2S]}) reflecting a natural phone pick-up delay, followed by a one-second pause later in the call (\texttt{[PEND1S]}).
A \texttt{[THINK]} near the end triggers S2 to look up real-time availability, with the result delivered as a single \texttt{\textless{}\ldots{}\textgreater{}} fragment after a brief holding phrase.

\begin{quote}
\small
\textbf{Assistant:} Hi, this is London Floral School. We noticed you're interested in flower design---just a quick intro to our flexible courses, from beginner to expert.

\textbf{User:} \texttt{[PEND2S]} Hey, I was just looking at flower shop pics. Does your school have programs for different experience levels?

\textbf{Assistant:} Our classes are fully flexible---book anytime. Whether you're a total beginner or looking to improve, we've got the right course. With 16 years of teaching experience, many graduates have opened their own flower shops.

\textbf{User:} Sounds good. Any special resources during the program?

\textbf{Assistant:} \texttt{\textless{}Fresh flower wholesale market on site; students practice with flowers imported directly from Dutch growers and auctions.\textgreater{}}Our biggest advantage is a live wholesale market on campus. Students pick fresh flowers for practice without going out early to buy them---all sourced directly from Dutch growers and auctions.

\textbf{User:} That's really convenient. I'm especially interested in wedding floral design. Do you teach actual wedding setups?

\textbf{Assistant:} Sure, our wedding floristry class is super popular. We cover bridal bouquets, table centerpieces, and trendy wedding arches---you'll learn structure and secure arrangement techniques.

\textbf{User:} \texttt{[PEND1S]} The flower arch sounds great, but I'm pretty busy. How long does this course take?

\textbf{Assistant:} Don't worry---we know time is precious. The wedding floristry course is modular; learn on weekends or evenings at your own pace.

\textbf{User:} Good. I'm only free weekends next month. Can you check if there are classes then? I'd like to decide soon.

\textbf{Assistant:} Sure, just a moment while I check next month's weekend schedule. \texttt{[THINK]}

\textbf{Assistant:} Sorry to keep you waiting. \texttt{\textless{}Weekend class sessions on the 12th and 19th next month still have availability.\textgreater{}}Got it---there are weekend classes on the 12th and 19th next month, and spots are still open. I can hold one for you and send the detailed outline to your phone shortly.
\end{quote}

\section{Scenario Distribution and Data Statistics}
\label{sec:appendix-stats}

\subsection{Interaction Pattern Distribution}

Each training sample is annotated by the Director with one or more temporal control tokens.
Because a single dialogue commonly exhibits multiple interaction phenomena---for example, a conversation may contain both an interruption and a shared silence---the categories below are \emph{not} mutually exclusive.
Table~\ref{tab:token-distribution} reports the proportion of training samples that contain each pattern.

\subsection{Scenario Seed Distribution}

Beyond token-level statistics, each sample is seeded with a primary interaction type that determines the overall dialogue structure.
Table~\ref{tab:scenario-distribution} shows the distribution of primary interaction types across the scenario seeds.

\subsection{Additional Data Statistics}

Table~\ref{tab:data-stats} summarizes key statistics of the annotated training corpus.

AI contributed to the polishing of this paper.

\end{document}